\documentstyle[aps,prl,epsf,twocolumn,epsfig,floats]{revtex}

\begin{document}

\wideabs{

\title{ On the detection of time-reversal symmetry breaking by
ARPES with circularly polarized light in Bi2212}

\author{N.P. Armitage and Jiangping Hu}

\address{ Department of Physics and Astronomy, University of California, Los
Angeles, CA 90095} \maketitle

\begin{abstract}

We argue that that in their recent experiment in which they claim
to have found evidence for a time-reversal symmetry broken state,
Kaminski \textit{et al.} overlooked small temperature dependent
changes in the superstructure of Bi2212.  These subtle changes may
manifest themselves by changing the final state configurations of
the photoemission process and thus invalidate their ultimate
conclusions.

\end{abstract}

}

Kaminski \textit{et al.} \cite{Kaminski} recently reported the
results of an experiment in which they found a small but
significant asymmetry ($\approx$3\%) in the photoemission
intensity of the high-T$_c$ superconductor
Bi$_2$Sr$_2$CaCu$_2$O$_{8+\delta}$ (Bi2212) below its pseudogap
temperature $T^*$ when using light of different helicities.   This
was interpreted as indicative of a hidden time reversal symmetry
(TRS) breaking phase in the underdoped regime.  However, as there
is ample evidence for low temperature structural changes in the
underdoped regime of Bi2212, we feel that Kaminski \textit{et al.}
are incorrect to infer the existence of a TRS breaking state from
their experiment.

The Kaminski \textit{et al.} experiment is not directly sensitive
to TRS and is, in fact, a probe of reflection symmetry (RS).
Although this point is made in the theoretical proposal
\cite{SimonVarma} and implicit in the experiment, it is worth
making explicit. It can be easily seen by considering the
photoemission optical matrix element

\begin{eqnarray}
M^r(\vec{k})=\sum_{fi}|<\psi_f|\hat{\Delta}^r|\psi_i>|^2,
\nonumber
\end{eqnarray}

where $\hat{\Delta}^{r(l)}$ is the dipole operator for right
(left) circularly polarized light and $|\psi_i>$ and $|\psi_f>$
are, for illustration purposes, one electron states.  With a
reflection symmetry operation $\hat{R}$, the terms of the above
matrix element equal
$|<\psi_f|\hat{R}^{-1}\hat{R}\hat{\Delta}^r\hat{R}^{-1}\hat{R}|\psi_i>|^2$.
If states $|\psi_i>$ and $|\psi_f>$ are eigenstates of reflection,
this quantity is proportional to $<\psi_f|\hat{\Delta}^l|\psi_i>$.
The associated matrix element can be defined as $M^l$ and
therefore the dichroism quantity probed by Kaminski \textit{et
al.} $D=M^r - M^l$ is sensitive to reflection.  However, because a
crystal's point group and reflection symmetries are only preserved
if all magnetic moments from orbital currents or spin are confined
to the irreducible unit cell, all proposed TRS breaking states for
the cuprate superconductors break RS across at least one mirror
plane.  In this regard, RS breaking is a necessary, but not
sufficient condition for TRS breaking.  In order for it to be a
sufficient there can be no structural features that break RS.  A
system that has a dichroism photoemission signal across a certain
direction and no structural aspects that break RS across that
direction may be a candidate for a TRS breaking state, but a state
that breaks RS alone can also give such an effect.

Although the $\Gamma \leftrightarrow (\pi,0)$ direction that
Kaminski \textit{et al.} use as a mirror plane in their experiment
is superficially a mirror plane of an idealized CuO$_2$ square
lattice,  RS is broken across this direction in Bi2212 by the well
known $\hat{b}$ direction [$\Gamma \leftrightarrow (\pi,\pi)$]
incommensurate superstructure modulation, which then ceases to be
a mirror plane \cite{Gao,Kirk}.  In principle, a supermodulation
at 45$^\circ$ to the $\Gamma \leftrightarrow (\pi,0)$ plane would
give a maximum dichroism signal at $(\pi,0)$.  As evidenced by the
experiment however, whatever the effects of the superstructure
are, they apparently do not cause an appreciable dichroism at
high-temperatures.  The conclusion that the low temperatures
effect is not caused by the superstructure and instead by a TRS
breaking state is only reasonable if there are no changes in the
modulation as a function of temperature and doping. If there are
such changes, it would seriously undermine the claim that that TRS
breaking could be inferred from RS breaking.  Monitoring only a
single main Bragg peak $via$ x-rays, as done by Kaminski
\textit{et al.}, is likely to be a very incomplete measure of
temperature dependent changes for a complicated oxygen-rich
incommensurate crystal like Bi2212.

We should note that although Kaminski \textit{et al.} argued from
their x-ray diffraction and the fact that their observed
dispersion is symmetric around $(\pi,0)$ that any temperature
dependent changes in the crystal structure were much smaller than
that necessary to cause a macroscopic shift of the mirror plane of
the near E$_F$ electronic states, this point is irrelevant.
Because $(\pi,0)$ is not in a true mirror plane of the Bi2212
crystal, any changes in the structure do not have to be such as to
influence the spectral function $A(\vec{k},\omega$) of the near
E$_F$ states greatly (of which the dispersion is indicative of)
greatly; the purported measurement is of the photoemission optical
matrix element $M^{r,l}$ which will be sensitive to subtle changes
in the hybridization of the final state configuration $|\psi_f>$.

There are a number of studies that do, in fact, suggest low
temperature structural changes of the incommensurate modulation
for underdoped samples.  For instance, Anderson \textit{et al.} in
a series of careful ultrasonic measurements found a sharp internal
friction loss peak, indicative of a bulk structural change, at 167
K \cite{ARAnderson} in only oxygen deficient (underdoped) Bi2212
samples.  Miles \textit{et al.} in a detailed Rietveld refinement
of neutron diffraction data found a sharp discontinuity in the
$\hat{b}$ axis lattice parameter as well as the superstructure
period for underdoped Bi2212 crystals around 160 K \cite{Miles2}.
Although the only comprehensive temperature dependent x-ray study
of the incommensurate structure found no such strong anomalies
(perhaps due to the weak oxygen sensitivity of x-rays, which makes
them a poor probe for this type of system), it did find a
suppression of the satellite intensities around $\sim$130 K
\cite{Johnson}.  The doping and temperature dependence of these
structural changes mimic that of T$^*$.

Generally speaking, strong temperature dependent interlayer
stresses arise in such low dimensional incommensurate structures
due to the differences in expansivities of intralayer bonds
\cite{Bak}.  Such stresses may be relieved by the vacancies in
oxygen deficient Bi-O layers through subtle
reconstructions\cite{Surface}.  As judged from the above
experiments which see changes in the 130-160 K range, such effects
may mimic a doping dependence which correlates with T$^*$.
Because the changes found in Bi2212 violate the structural
symmetry condition necessary to infer a TRS breaking state from RS
breaking, we feel that Kaminski \textit{et al.} cannot reasonably
deduce the existence of an exotic TRS violating state from their
experiment.



\begin{thebibliography}{1}


\bibitem{Kaminski} A. Kaminski \textit{et al.}, Nature \textbf{416}, 610 (2002).

\bibitem{SimonVarma}  M.E. Simon and C.M. Varma, Phys. Rev. Lett. \textbf{89}, 247003 (2002).

\bibitem{Gao} Y. Gao \textit{et al.}, Science \textbf{241}, 954 (1988).

\bibitem{Kirk} M.D. Kirk \textit{et al.}, Science \textbf{242}, 1673(1988).

\bibitem{ARAnderson} A.R. Anderson \textit{et al.}, Physica C \textbf{281}, 356 (1997);  A.R. Anderson \textit{et al.}, Supercond. Sci. Technol. \textbf{5}, 258 (1992).

\bibitem{Miles2}  P.A. Miles \textit{et al.}, Phys. Rev. B \textbf{55}, 14632 (1997).

\bibitem{Johnson} S.T. Johnson and P.D. Hatton, Sol. St. Comm. \textbf{94}, 261 (1995).

\bibitem{Bak}  P. Bak, Rep. Prog. Phys. \textbf{45}, 587 (1982).

\bibitem{Surface}  Such stresses can also be relieved at the
surface.  As ARPES is an eminently surface sensitive probe, the
surface structural symmetry needs to be considered separately.
Although LEED studies have claimed that the surface structure is
commensurate with the bulk, analysis was not performed at the
requisite level of detail to allow strong statements about subtle
differences \cite{Lindberg}.  In fact, STM routinely shows
\cite{Kirk,Inoue,SHPan} a strong additional surface modulation of
the bismuth atoms at approximately the superstructure periodicity
of which no corresponding feature has been observed in bulk
sensitive x-ray diffraction.  The differences between surface and
bulk structure means that this issue needs to be investigated in
more detail before one can rule out doping dependent low
temperature surface reconstructions.

\bibitem{Lindberg}  P.A.P. Lindberg \textit{et al.}, App. Phys. Lett. \textbf{53}, 2563 (1988); R. Claessen, Phys. Rev. B \textbf{39}, 7316 (1989).

\bibitem{Kirk} M.D. Kirk \textit{et al.}, Science \textbf{242}, 1673(1988).

\bibitem{Inoue} A. Inoue \textit{et al.}, Physica C \textbf{233}, 49 (1996).

\bibitem{SHPan} S.H. Pan \textit{et al.}, App. Phys. Lett. \textbf{73}, 58 (1998).


\end{thebibliography}
\end{document}